\documentstyle[11pt,epsfig,epsf,axocolor]{article}
\newcommand{\ba}{\begin{array}}
\newcommand{\ea}{\end{array}}
\newcommand{\bd}{\begin{displaymath}}
\newcommand{\ed}{\end{displaymath}}
\newcommand{\be}{\begin{equation}}
\newcommand{\ee}{\end{equation}}
\newcommand{\bea}{\begin{eqnarray}}
\newcommand{\eea}{\end{eqnarray}}

\def\q2 {q^2}

\begin{document}
\begin{center}
{\Large\bf  Leptoquark Signals via $\nu $ interactions at Neutrino Factories}
\\[15mm]
\vskip 15pt
{\sf Poonam Mehta $^{a, \,\!\!}$
\footnote{E-mail address: pm@ducos.ernet.in, pmehta@physics.du.ac.in }}, 
{\sf Sukanta Dutta $^{b, \,\!\!}$
\footnote{E-mail address: Sukanta.Dutta@cern.ch}}
and  
{\sf Ashok Goyal $^{a,c, \,\!\!}$
\footnote{E-mail address: agoyal@ducos.ernet.in, goyal@iucaa.ernet.in }}
\vskip 8pt
$^a${\em Department of Physics \& Astrophysics, University of Delhi, Delhi 110 007, India }\\

$^b${\em Physics Department, S.G.T.B. Khalsa College, University of
Delhi, Delhi 110 007, India }\\

$^c${\em Inter University Center for Astronomy \& Astrophysics, Ganeshkhind, Pune 411 007, India }\\

\end{center}
\vskip 5pt 
\begin{abstract}
  
  The accurate prediction of neutrino beam produced in muon decays 
  and the absence of opposite helicity contamination for a particular 
  neutrino flavour makes a future neutrino factory (NF) based on a muon 
  storage ring (MSR), the ideal place to look for the lepton flavour 
  violating (LFV) effects.
  In this letter, we address the contribution  of mediating LFV 
  leptoquarks (LQ) in $\nu (\bar\nu)-{\rm N}$ interactions 
  leading to production of $\tau$'s and wrong sign $\mu$'s at MSR 
  and investigate the region where LQ interactions are significant 
  in the near-site and short baseline experiments.
\end{abstract}
\vskip 0.5 cm
\setcounter{footnote}{0}
\begin{section}{Introduction}          
Recent results from Super Kamiokande and other experiments
\cite{superk_results} strongly suggest $\nu_\mu$-$\nu_\tau$ 
oscillation as the dominant oscillation mode, 
in order to explain the atmospheric $\nu_\mu$ 
deficit. Similarly, the results for solar neutrino problem point 
towards $\nu_e$-$\nu_\mu$ oscillation as the favoured solution \cite{solar}.
In fact, the prime goal of next generation neutrino physics experimental 
studies ( for eg. NF based on MSR ) is to explore the physics beyond 
SM to unfold the mystery of the neutrino mass hierarchy and 
confirm the nature of neutrino flavour conversion \cite{storage_ring}. 
At MSR with a $\mu^-$ ($\mu^+$) beam, roughly $\simeq 10^{20}$ 
muons are allowed to decay per year giving rise to nearly
equal number of $\nu_\mu$ ($\bar \nu_\mu$) and $\bar \nu_e$ ($\nu_e$).
These $\nu $ ($\bar\nu$)'s at the detector,  may or may not have 
changed their flavour due to oscillation of neutrino mass eigenstates, 
which on interaction with matter produce associated charged 
leptons \cite{tau_appearance}. 
However, there can be effective LFV interactions 
motivated from new physics  which may give rise to 
charged leptons in the final state as expected 
through $\nu$($\bar \nu$)-oscillations \cite{rparity}.

In this backdrop, it is worthwhile to study the production of 
$\tau$ and wrong sign $\mu$ via LQ as mediators 
which occur naturally in Grand Unified Theories, 
Superstring inspired $E_6$ models and in Technicolor models \cite{lqs}. 
There have been numerous phenomenological studies to put constraints 
on LQ from low energy flavour changing neutral current (FCNC) 
processes which are generated by both the scalar and vector LQ 
interactions, since there is no reason why the quark-lepton couplings 
with LQ have to be simultaneously diagonal in quark and lepton mass 
matrices. Direct experimental searches for leptoquarks have also been 
carried out at the e~p collider and bounds obtained \cite{davidson,lqbounds}.
In this letter, we compute and analyse the contribution of mediating LFV LQ 
in $\nu $($\bar \nu $)-$N$ charged current (CC) interactions, 
including constraints obtained from low energy phenomenology.

The most general expression for the event rate per kilo Ton (kT) 
of the target per year for any charged lepton flavour
$l_k$, obtained via CC interaction of $\nu_j$ beam{\footnote 
{ $k=j$ for the  SM Lepton Flavour Conserving 
situation}}  produced  as a result of oscillation from an 
initial $\nu_i$ beam can  be written as :
\begin{equation}
{\cal N}_{l_k^-,l_k^+} = {\cal N}_n \int { 
{{ d^2 \sigma^{\nu,{\bar \nu}}} \left(\nu_j(\bar \nu_j) q 
\longrightarrow l_k^-(l_k^+) q'\right) \over {dx~dy} }}
\left[{d N_{\nu,\bar \nu} \over d E_{\nu_i,\bar \nu_i}}\right] 
{\cal P}_{osc} (\nu_i(\bar \nu_i) \longrightarrow \nu_j(\bar \nu_j)) 
d E_{\nu_i(\bar \nu_i)}\,q(x)\,~dx~dy
\end{equation}
\noindent where, ${\cal N}_n$ is the number of nucleons present kT 
of the target material, x and y are the Bjorken scaling variables, q and q' 
are the quarks in the initial and final states, respectively 
and ${\cal P}_{osc}$ is the oscillation probability. 
The differential parton level cross-section $
{{ d^2 \sigma^{\nu,{\bar \nu}}} \over  {d x~ d y} }$ is 
$\left[{ \left\vert{\cal M}(x,y)\right\vert ^2 \over {32 \pi \hat s} }\right] 
\left[2 \lambda^{-1/2}(1,0,{m_{l}^2 \over \hat s} )\right]$
where , $\hat s$ is the parton level CM energy, 
$m_l$ is the mass of the final-state lepton and
$\lambda^{1/2}(x,y,z) = x^2 + y^2 +z^2 -2xy -2xz -2 yz $ 
is the Michael parameter 
and $\left[{d N_{\nu,\bar \nu} \over d E_{\nu_i,\bar \nu_i}}\right]$ is the differential $\nu$ ($\bar \nu$) flux. 
For the two flavour oscillation scenario{\footnote {For the present case, 
it is sufficient to illustrate the main ideas by considering only the
two flavour oscillations in vacuum.}}, the  probability 
${\cal P}_{osc} (\nu_{i}\rightarrow \nu_j)$ 
is $ \sin^2 2\theta_m\,\, \sin^2\left[ 
1.27\, \Delta m^2[eV^2]\, {L[km]\over E_\nu[GeV]}\right]$, 
where, $L$ is the baseline length, $E_\nu$ is the neutrino 
energy, $\Delta m^2$ is the mass-squared difference between 
the corresponding physical states, and $\theta_m$ is mixing angle 
between flavours. 
Here $q(x)$ is the quark distribution function. The general characteristics 
of $\tau$  and wrong sign $\mu$ production in the  oscillation 
scenario (OS), for example are given by Dutta {\em et al.} 
\cite{tau_appearance}.
\par The effective Lagrangian with the most general dimensionless, 
$SU(3)_c$X$SU(2)_L$X$U(1)_Y$ invariant couplings of {\it scalar} and {\it 
vector} LQ satisfying baryon ($B$) and lepton number ($L$) conservation 
(suppressing colour, weak isospin and generation (flavour) indices )
is given \cite{lq_lag} by:
\bea
{\cal {L}} &=& {{\cal {L}}_{\vert F\vert =2}} + {{\cal {L}}_{\vert F\vert =0}}\,\,\,\,\, \,\,\,\, {\rm where} \nonumber\\
{{\cal {L}}_{\vert F\vert =2}} &=& \left[ g_{1L}\, \bar q^c _L\, i\, \tau_2 \,l_L +\, g_{1R} 
\,\bar u^c_R \,e_R \right] \,S_1 +\,\tilde g_{1R}\, \bar d^c_R \,e_R \,\tilde S_1 
+\,g_{3L}\,\bar q^c_L \,i \,\tau_2 \,{\vec \tau} \,l_L \,{\vec S}_3\nonumber \\ 
&+& \,\left[ g_{2L} \,\bar d^c _R \,\gamma^\mu \,l_L + \,g_{2R} \,\bar q^c_L\, 
\,\gamma^\mu \,e_R \right] \,V_{2\mu} + \,\tilde g_{2L} \,\bar u^c_R \,\gamma^\mu \,l_L\,\tilde V_{2\mu} + \,{\rm \bf c.c.},\nonumber \\
{{\cal {L}}_{\vert F\vert=0}} &=& \left[ h_{2L} \,\bar u_R \,l_L +\, h_{2R} 
\,\bar q _L\, i \,\tau_2 \,e_R \right] \,R_2 + \,\tilde h_{2L} \,\bar d _R \,l_L 
\,\tilde R_2 + \,\tilde h_{1R} \,\bar u _R \,\gamma^\mu \,e_R \,\tilde U_{1\mu}
 \nonumber \\
&+&\left[ h_{1L} \,\bar q _L \,\gamma^\mu \,l_L +
\,h_{1R} \,\bar d _R \,\gamma^\mu \,e_R \right] \,U_{1\mu} 
+ h_{3L} \,\bar q _L \,{\vec \tau} \,\gamma^\mu \,l_L \,U_{3\mu} + {\rm \bf c.c.}
\eea
\noindent where  $q_L$, $l_L$ are the left-handed quarks 
and lepton doublets and $e_R$, $d_R$, $u_R$ are the right-handed 
charged leptons, down- and up-quark singlets respectively . 
The Scalar (i.e. $S_1$, $\tilde S_1$, {\bf $S_3$}) and 
Vector (i.e. $V_2$, $\tilde V_2$) LQ carry fermion number 
${\rm F}=3{\rm B}+{\rm L}=-2$, 
while the Scalar (i.e. $R_2$, $\tilde R_2$ ) and Vector (i.e. $U_1$, 
$\tilde U_1$, {\bf $U_3$}) LQ have ${\rm F}=0$.
Using this Lagrangian we discuss below the production of $\tau$'s 
and wrong sign $\mu$'s along with the standard Mass-Mixing solution 
of neutrino oscillation case.
\end{section}

\begin{section}{Tau Appearance at a NF}
\begin{figure}[h]
\begin{picture}(280,100)(0,0)
\vspace*{- 1in}
\ArrowLine(60,-5)(100,30){psRed}
\Text(60,-16)[r]{$\nu_\mu$}
\ArrowLine(100,30)(57,65){psRed}
\Text(60,80)[r]{${u,\, c}$}
\DashLine(100,30)(170,30){4}{psBlue}
\Text(135,40)[b]{$R,\,U$}
\ArrowLine(170,30)(200,70){psRed}
\Text(202,80)[b]{${\tau^-}$}
\ArrowLine(210,-3)(170,30){psRed}
\Text(202,-16)[b]{$d$}
\Text(135,-15)[b]{(a)}

\ArrowLine(265,-5)(305,30){psRed}
\Text(265,-16)[r]{$\nu_\mu$}
\ArrowLine(262,65)(305,30){psRed}
\Text(265,80)[r]{$d$}
\DashLine(305,30)(375,30){4}{psBlue}
\Text(340,40)[b]{$S,\,V$}
\ArrowLine(375,30)(405,70){psRed}
\Text(402,76)[b]{$\tau^-$}
\ArrowLine(375,30)(415,-3){psRed}
\Text(407,-16)[b]{$u,\, c$}
\Text(330,-15)[b]{(b)}
\end{picture}
\vskip .3in
\caption{{\em  $\tau^-$ from scalar \& vector LQ: (a) u-channel 
process corresponding to $\vert {\rm F}\vert$=0 LQ and 
(b) s-channel process corresponding to $\vert{\rm }F\vert $=2 LQ.}}
\label{fig:dia1}
\end{figure}
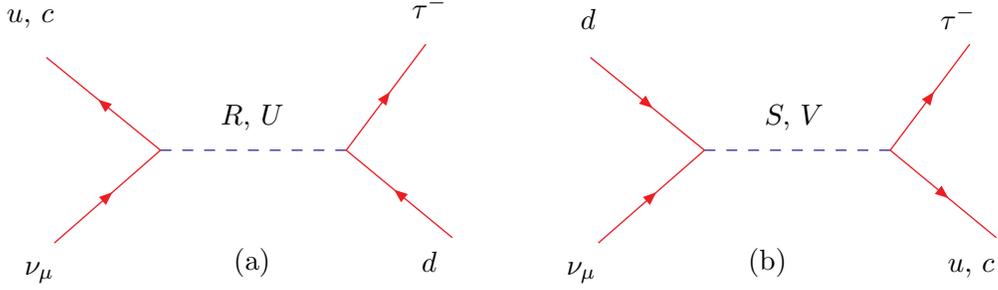
We consider the production of $\tau^-$ from 
unoscillated $\nu_\mu$ (obtained from $\mu^-$ decay) 
through LFV interactions with nucleon via u-channel 
processes for $\vert {\rm F}\vert=0$ case  (figure \ref{fig:dia1}(a))
and s-channel processes for $\vert {\rm F}\vert=2$ 
case  (figure \ref{fig:dia1}(b)) LQ unlike OS 
where $\tau^-$ are produced through interaction of $\nu_\tau$ 
( oscillated from $\nu_\mu$ with $\Delta m^2 = 0.0023$ eV$^2$ and 
$\sin^2 (2\theta_m) = 1.0$ ) with the nucleon.
There are four processes contributing to $\tau^-$ production in 
the u-channel (figure \ref{fig:dia1}(a)), 
one mediated by the charge $=2/3$, scalar LQ $(R_2)$ with $T_3=-1/2$ 
and three by the vector LQ $( U_{1\mu}, U_{1\mu}, U_{3\mu} )$ with 
$T_3=0$ each 
{\footnote {In our notation, $R_2^a$ denotes $R_2$ with $T_3=-1/2$ 
and $U_{3\mu}^0$ implies $U_{3\mu}$ with $T_3=0$.}}, 
where $T_3$ is the weak isospin. 
The matrix element squared for all the u-channel processes is
\bea
{ \left\vert{\cal M}_{LQ}^{u-chann}(\nu_\mu d \longrightarrow \tau^- u)
\right\vert ^2 } = \left[ \hat u (\hat u - m_\tau ^2)\right] 
\Biggl[{ \left\vert h_{2L} \, h_{2R}\right\vert ^2 \over (\hat u - M_{ R _{2} ^a }^2)^2 }\Biggr]
~+~ \left[4 \hat s (\hat s - m_\tau ^2)\right] 
\Biggl[{\left\vert h_{1L}\, \right\vert^4 \over (\hat u - M_{U_{1\mu}}^2)^2}\nonumber \\ \nonumber \\
~+~ { \left\vert h_{3L}\right\vert^4 \over (\hat u - M_{U _{3\mu} ^0}^2)^2} 
~-~ 2 { \left\vert h_{1L} \, h_{3L} \right\vert^2 \over (\hat u - M_{U_{1\mu}}^2)  
(\hat u - M_{U_{3\mu}^0}^2) } \Biggr]
~+~ \left[ 4\hat t (\hat t - m_\tau ^2)\right] 
\Biggl[{ \left\vert h_{1L}\, h_{1R} \right\vert^2 \over (\hat u - M_{U_{1\mu}}^2)^2} \Biggr] 
\eea
\\
\noindent where, the Mandelstam variables at the parton level are given by 
${\hat s}=(p_{\nu_\mu}+p_d)^2$, ${\hat t}=(p_{\nu_\mu}-p_{\tau^-})^2$ 
and 
${\hat u}=(p_{\nu_\mu}-p_{u,c})^2$, with $p_i$ denoting the four 
momemtum of the $i^{th}$ particle. 
In the s-channel, three processes are mediated by charge $=-1/3$, 
scalar LQ $( S_1,S_1,S_3 )$ with $T_3=0$, while the fourth one is mediated 
by a vector LQ $(V_2)$ with $T_3=-1/2$ (figure \ref{fig:dia1}(b))
{\footnote {In our notation, $S_3^0$ denotes $S_3$ with $T_3=0$ and 
$V_{2\mu}^a$ implies $V_{2\mu}$ with $T_3=-1/2$.}}.
The matrix element squared for s-channel processes is
\bea
{ \left\vert{\cal M}_{LQ}^{s-chann}(\nu_\mu d \longrightarrow \tau^- u)\right\vert ^2 } =
\left[\hat s (\hat s - m_\tau ^2)\right] \Biggl[ { \left\vert g_{1L}\right\vert ^4 \over 
(\hat s - M_{S_1}^2)^2} ~+~ 
{ \left\vert g_{1L} \, g_{1R} \right\vert^2 \over (\hat s - M_{S_1}^2)^2}  ~+~
{\left\vert  g_{3L}\right\vert^4 \over (\hat s - M_{S_{3}^0}^2)^2} \nonumber \\ \nonumber \\
~-~ 2 {\left\vert g_{1L} \, g_{3L}\right\vert^2 \over (\hat s - M_{S_1}^2)  
(\hat s - M_{S_{3}^0}^2) }\Biggr] ~+~ 
\left[ 4\hat t (\hat t - m_\tau ^2)\right] 
\Biggl[{ \left\vert g_{2L}\,  g_{2R}\right\vert^2 \over (\hat s - M_{ V _{2\mu} ^a}^2)^2 } \Biggr]
\eea
\\

\par In order to demonstrate the  behaviour of the $\tau$ 
production rate, we consider the contribution from LQ carrying 
different fermion numbers separately, which implies that {\em either} the 
$h's$ {\em or} the $g's$ ( contributing to a given process ) are non-zero 
at a time. For simplicity, we have taken the masses of scalar 
and vector LQ and couplings $h's$ ( $g's$ ) 
for $\vert {\rm F}\vert=0$ ( $\vert {\rm F}\vert=2$ ) to be equal. 
We have used CTEQ4LQ parton distribution functions \cite{cteq} 
to compute the event rates.   
To study the variation of $\tau$ events w.r.t $E_\mu$ and baseline length $L$, 
we have plotted the events for two different LQ masses 250 GeV \& 500 GeV 
respectively using the product of couplings to be equal to 
$\alpha_{em}$. It should also be noted that since there are no strong bounds 
on the LQ interacting with a charm quark and a $\tau^-$ lepton
existing in the literature, the cross-section for c\,$\tau^-$ production
in the $\nu_\mu$\,N DIS is governed by the flavour violating couplings
between the second and third generation, which are not restricted by the 
bounds from the rare decays. The problem of charm detection and elimination 
of possible backgrounds however, needs to be tackled before the 
large available area in the parameter space can be explored.
\begin{figure}[h]
\vskip -4cm
\centerline{ 
\epsfxsize=12cm\epsfysize=24cm\epsfbox{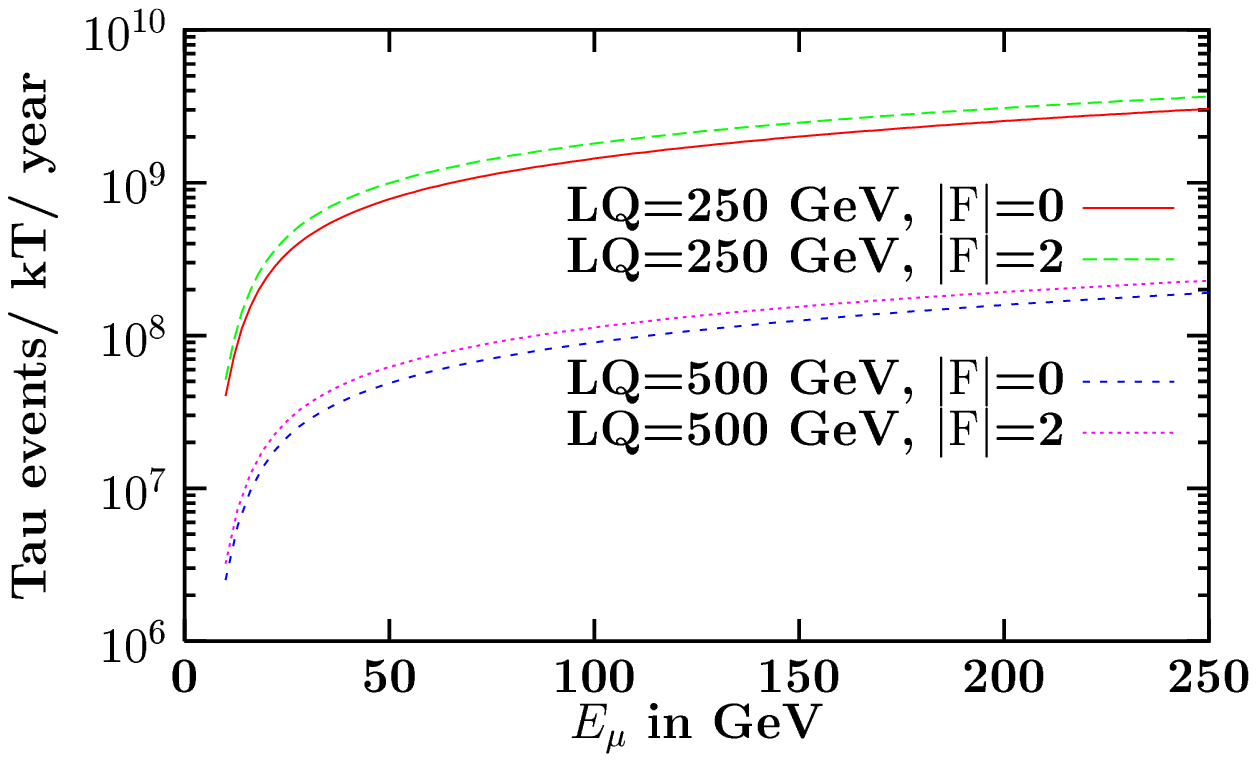}
	\hspace*{-3.5 cm}
\epsfxsize=12cm\epsfysize=24cm
                     \epsfbox{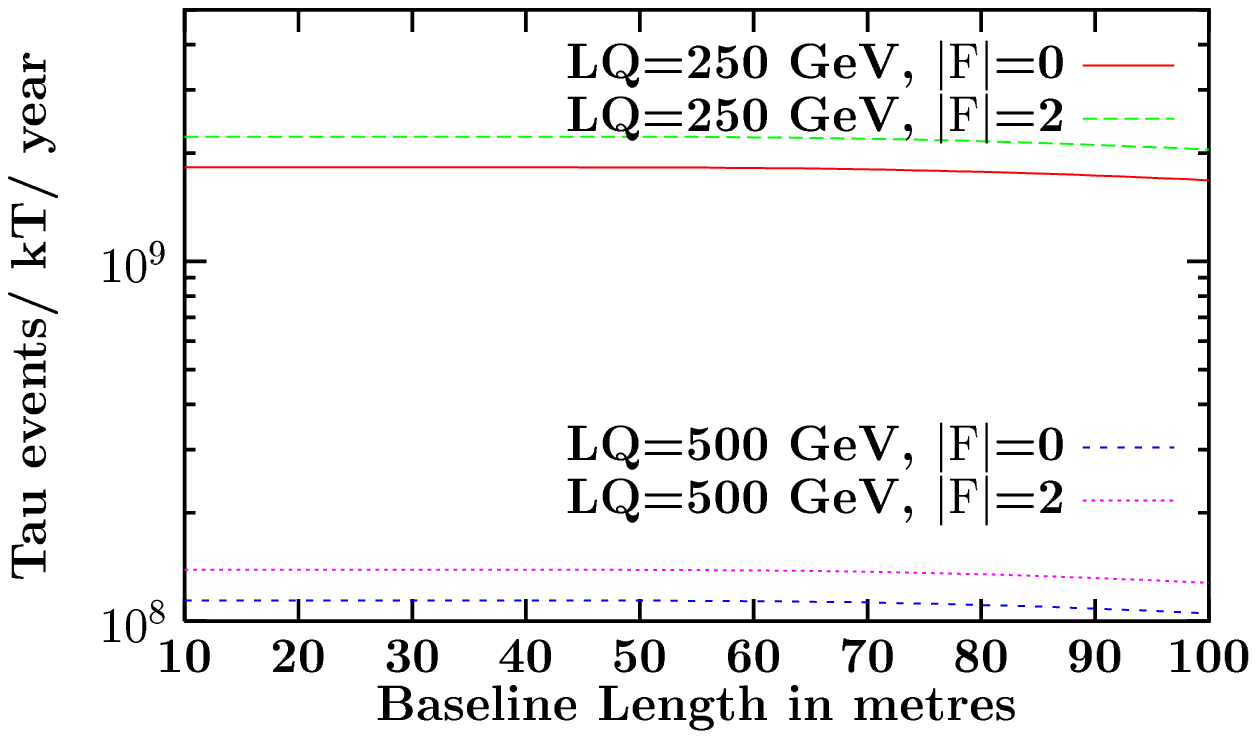}
}
\vskip -13cm
\hskip 4cm
{\bf (a)}
\hskip 8cm
{\bf (b)}
\vskip -3.5cm
\centerline{ 
\epsfxsize=12cm\epsfysize=24cm\epsfbox{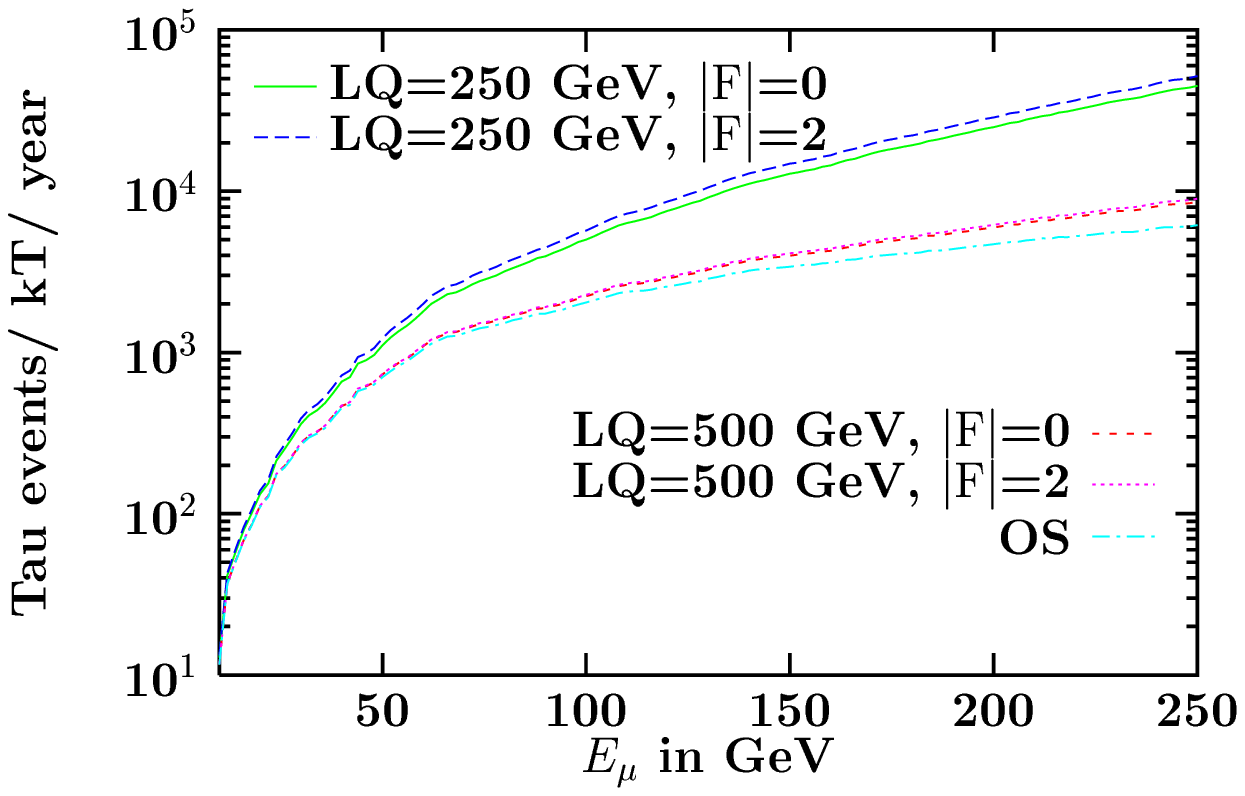}
	\hspace*{-3.5 cm}
\epsfxsize=12cm\epsfysize=24cm
                     \epsfbox{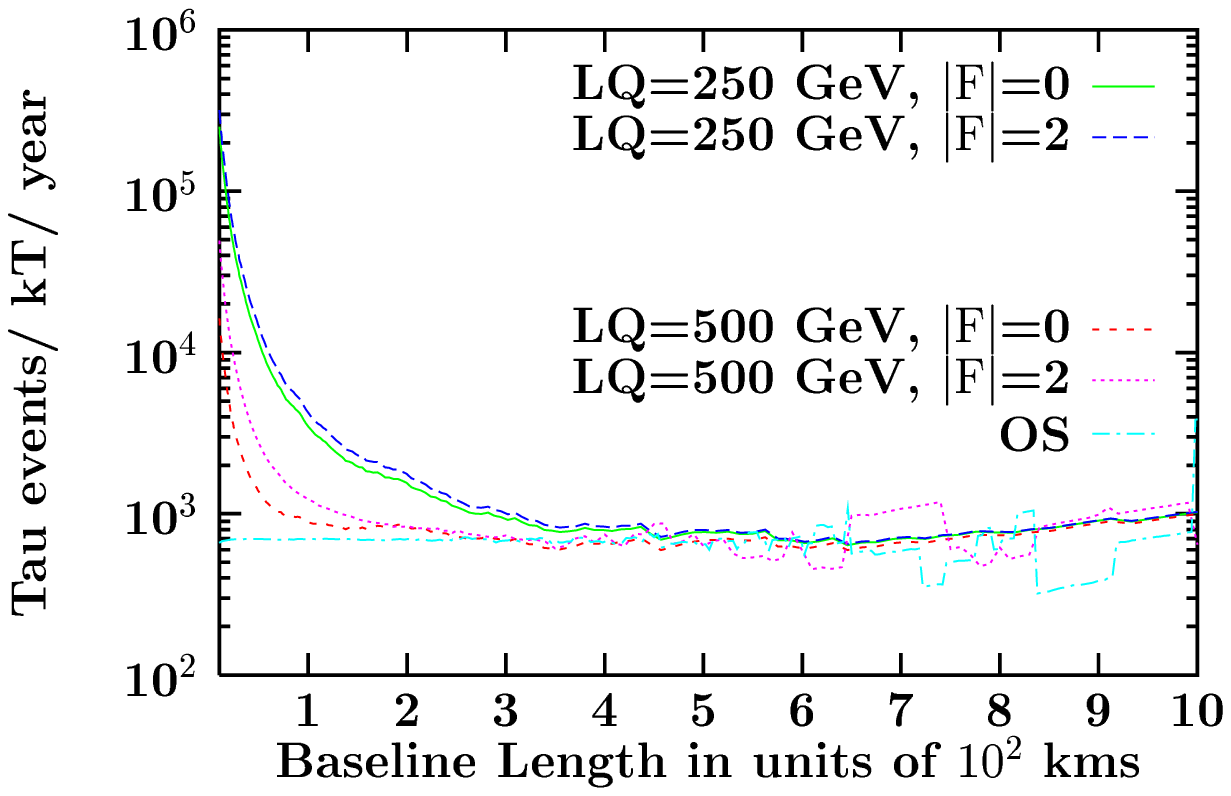}
}
\vskip -13cm
\hskip 4cm
{\bf (c)}
\hskip 8cm
{\bf (d)}
\caption{{\em Variation $\tau$-events ( from osc. and LQ ) for a 1kT detector,
 LQ mass 250 \& 500 GeV and product of LQ coulings = 0.089 with : 
(a) muon beam energy for a baseline length 40 meters 
and sample detector area 0.025 m$^2$, 
(b) baseline length for muon beam energy 50 GeV and detector area 0.025 m$^2$,
(c) muon beam energy for a baseline length 250 kms and sample detector 
area 100 m$^2$,  
(d) baseline length for muon beam energy 50 GeV and detector area 100 m$^2$.
}}
\label{fig:fig2}
\end{figure}
In figure \ref{fig:fig2}(a), we plot the net contribution 
( from LQ and oscillation ) to tau events 
for a near-site experimental set-up w.r.t $E_\mu$. 
We have considered a detector with a sample area 
of .025 \,$m^2$ \cite{detector} and placed at 
40 \,$mts$ from the storage ring. 
It is worthwhile to mention that the contribution is predominantly 
from LQ as the oscillation is suppressed at such baseline length. 
We give similar curves in figure \ref{fig:fig2}(c)
with the detector placed at a baseline length of 
$250$ \,$kms$ ( K2K Proposal, {\it from KEK to Kamioka} ) and sample 
detector area of $100$ \,$m^2$ \cite{detector}.
Here the contribution of LQ is comparable to that of the oscillation. 
Figure \ref{fig:fig2}(b) shows the variation of events w.r.t. the
baseline length, $1$\,$m$ to $100$\, $m$ 
(appropriate for near-site experiment) for $E_\mu$ fixed at $50$ \,$GeV$. 
The graph clearly shows the independence of the tau events
with baseline length in this range, while in figure \ref{fig:fig2}(d) 
the behaviour of tau event rate is markedly different for 
short and medium baselines ($1 - 1000$ \,$kms$). 
Here, the LQ event rate falls off as $1/L^2$ to zero 
and hence the combined event rate for $\tau$ essentially merges 
with that due to oscillation alone.

The background for the signal of $\tau$ and the ways to 
eliminate them have been already discussed in detail in the existing 
literature ( see for example, reference \cite{cuts} ) and it is found out that
the missing-$p_T$ and isolation cuts taken together can remove the entire 
set of backgrounds due to charmed particle production, from 
unoscillated CC events and from the neutral current background. 
Recently, 
there have been theories that propose the existence of an extra neutral boson 
in many extensions of SM which lead to $\nu_\mu$ associated 
charm production \cite{charm}, which also acts as a source of background 
and need to be eliminated as far as detection of $\tau$ events 
are concerned.
The $\tau$-detection efficiency factor of 30\% \cite{rparity,detector,cuts} 
taken in the present calculation, adequately accounts for all 
the selection cuts ( including the cuts for missing $p_T$, 
isolation cut and the branching ratio ) required to eliminate 
the backgrounds. 

\noindent {\bf {Sensitivity Limits \,:}}
An estimate of the sensitivity limits on product of couplings and 
LQ masses can be based on the total number of events. 
Here we determine the range of LQ masses and product of LFV couplings, 
for which the number of signal events is equal to two and five times 
the square root of the OS events. 
Accepting this requirement of $2\sigma$ and $5\sigma$ effect as a 
sensible discovery criterion, 
we plot the corresponding contours in figure \ref{fig:fig3} 
for baseline length=40 m. 
Thus, non compliance of these estimate with experimental observation 
would mean that the lower region enclosed by the curve are ruled out 
at $2\sigma$ and $5\sigma$ level, respectively.
\begin{figure}[h]
\input{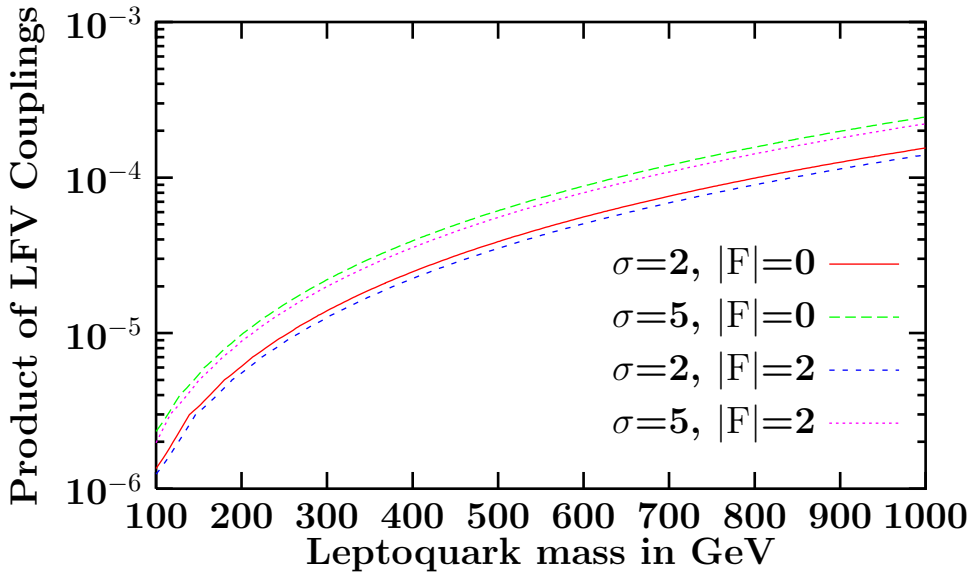}
\caption{{\em Contours for 2$\sigma$ and 5$\sigma$ effect 
for $E_\mu$=50 GeV, baseline length=40 meters and 
sample detector of area 2500 cm$^2$ and mass 1kT.}}
\label{fig:fig3}
\end{figure}
\end{section}
\begin{section}{Appearance of Wrong Sign Muons at a NF}
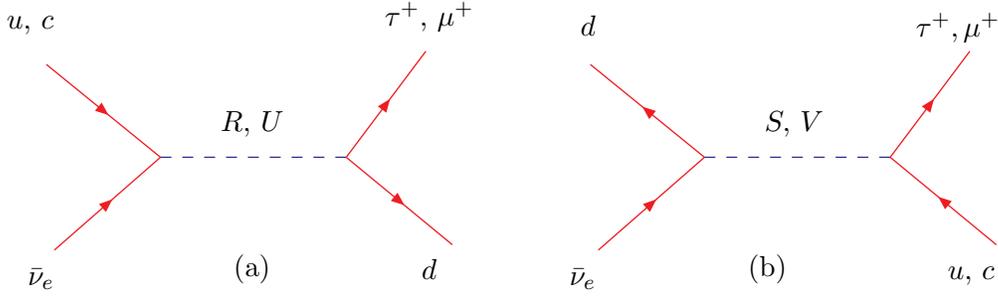
\begin{figure}[h]
\begin{picture}(280,100)(0,0)
\vspace*{- 1in}
\ArrowLine(60,-5)(100,30){psRed}
\Text(60,-16)[r]{$\bar\nu_e$}
\ArrowLine(57,65)(100,30){psRed}
\Text(60,80)[r]{${u,\,c}$}
\DashLine(100,30)(170,30){4}{psBlue}
\Text(135,40)[b]{$R,\,U$}
\ArrowLine(170,30)(200,70){psRed}
\Text(202,80)[b]{${\tau^+,\,\mu^+}$}
\ArrowLine(170,30)(210,-3){psRed}
\Text(202,-16)[b]{$d$}
\Text(135,-15)[b]{(a)}
\ArrowLine(265,-5)(305,30){psRed}
\Text(265,-16)[r]{$\bar\nu_e$}
\ArrowLine(305,30)(262,65){psRed}
\Text(265,80)[r]{$d$}
\DashLine(305,30)(375,30){4}{psBlue}
\Text(340,40)[b]{$S,\,V$}
\ArrowLine(375,30)(405,70){psRed}
\Text(402,76)[b]{$\tau^+,\mu^+$}
\ArrowLine(415,-3)(375,30){psRed}
\Text(407,-16)[b]{${u,\,c}$}
\Text(330,-15)[b]{(b)}
\end{picture}
\vskip .3in
\caption{{\em Production of $\tau^+\&\, \mu^+$ from scalar \& vector LQ: 
(a) s-channel process corresponding to $\left\vert {\rm F} \right\vert=0$ 
LQ and 
(b) u-channel process corresponding to $\left\vert {\rm F} \right\vert=2$ LQ.}}
\label{fig:dia2}
\end{figure}
In the OS, $\bar\nu_e$ from the parent $\mu^-$ beam can oscillate to 
either $\bar \nu_\mu$ or to $\bar \nu_\tau$ which give rise 
to $\mu^+$ and $\tau^+$, respectively. The $\tau^+$  
furthur decay muonically ($BR=17\%$ \cite{pdg}) and thus contribute to 
the $\mu^+$ events. 
However, it is worthwhile to mention here that one can hardly expect 
any $\mu^+$ events from oscillations since the neutrino mass-splitting 
required for the Mikheyev-Smirnov-Wolfenstein (MSW) solution to the 
solar neutrino problem \cite{solar} with matter-enhanced 
$\nu_e$-$\nu_\mu$ oscillation is $\Delta m^2 \simeq 10^{-5} eV^2$.
The situation is even worse for the case of Vacuum Oscillation 
solution which requires $\Delta m^2 \simeq 10^{-10} eV^2$. 
For the $\nu_e$-$\nu_\tau$ oscillation, there exists no experimental
support and so, the region of parameter space to be explored for such 
oscillation mode is not known at all. Thus, a significant event rate for 
wrong sign muons cannot be attributed to $\nu$-oscillation effects alone.

\par Here, we consider the production of $\mu^+$ from parent  
$\mu^-$ beam via {\em unoscillated} $\bar\nu_e$ through LFV interactions 
with nucleon mediated by LQ in two different ways:
(i) Direct Production of  $\mu^+$ as well as 
(ii) Production of $\tau^+$, which further decays leptonically to $\mu^+$. 
Both of these involve  s-channel processes corresponding to 
${\vert {\rm F}\vert}=0$ \& charge $=2/3$ (figure \ref{fig:dia2}(a)) LQ and 
u-channel processes corresponding to $\vert {\rm F}\vert=2$ 
\& charge $=-1/3$ (figure \ref{fig:dia2}(b)) LQ.
In figure \ref{fig:dia2}(a) out of four s-channel diagrams, 
one is mediated by a scalar LQ $( R_{2}^a )$ with $T_3=-1/2$, 
while the other three are mediated by vector 
LQ $( U_{1\mu}, U_{1\mu}, U_{3\mu}^0 )$ with $T_3=0$ each. 
The matrix element squared for all the four s-channel processes is 
\bea
{ \left\vert{\cal M}_{LQ}^{s-chann}(\bar {\nu_e} u \longrightarrow \mu^+ d)
\right\vert ^2 } = 
\left[\hat s (\hat s - m_\mu ^2)\right] 
\Biggl[{\left\vert h_{2L}\, h_{2R}\right\vert^2 \over (\hat s - M_{R_{2} ^a }^2)^2 }\Biggr] ~+~
\left[4(\hat s + \hat t)(\hat s +\hat t - m_\mu ^2)\right] 
\Biggl[ {\left\vert h_{1L}\right\vert^4 \over (\hat s - M_{U_{1\mu}}^2)^2} \nonumber \\ \nonumber \\
~+~  {\left\vert  h_{3L}\right\vert^4 \over (\hat s - M_{U_{3\mu}^0}^2)^2} 
~-~ 2  { \left\vert h_{1L} \, h_{3L}\right\vert^2 \over (\hat s - M_{U_{1\mu}}^2)  
(\hat s - M_{U_{3\mu}^0}^2) } \Biggr]
~+~ \left[4\hat t(\hat t - m_\mu ^2)\right]
\left[ { \left\vert h_{1L}\, h_{1R}\right\vert^2 \over (\hat s - M_{U_{1\mu}}^2)^2 }\right]
\label{wrs}
\eea
\noindent where, ${\hat s}=(p_{\bar {\nu_e}}+p_{u,c})^2$, ${\hat t}=
(p_{\bar {\nu_e}}-p_{\mu^+})^2$ 
and ${\hat u}=(p_{\bar {\nu_e}}-p_d)^2$, with $p_i$ 
denoting the four momemtum of the $i^{th}$ particle. 
In figure \ref{fig:dia2}(b) out of four u-channel diagrams   
three are mediated by scalar LQ $( S_1, S_1, S_3^0 )$ 
with $T_3=0$ each and the fourth diagram is mediated by a 
vector LQ $( V_{2\mu}^a )$ with $T_3=-1/2$. 
The matrix element squared for all the four u-channel processes corresponding 
to $\vert {\rm F}\vert=2$ is 
\bea
{ \left\vert{\cal M}_{LQ}^{u-chann}(\bar {\nu_e} u \longrightarrow \mu^+ d)
\right\vert ^2 } = \left[ \hat u (\hat u - m_\mu ^2)\right] 
\Biggl[ { \left\vert g_{1L}\right\vert^4 \over (\hat u - M_{S_1}^2)^2} ~+~ 
{ \left\vert g_{1L} \, g_{1R}\right\vert^2 \over (\hat u - M_{S_1}^2)^2}  ~+~
{\left\vert  g_{3L}\right\vert^4 \over (\hat u - M_{S_{3}^0}^2)^2} \nonumber \\ \nonumber \\
~-~ 2 { \left\vert g_{1L} \, g_{3L}\right\vert^2 \over (\hat u - M_{S_1}^2)  
(\hat u - M_{S_{3}^0}^2) }\Biggr] ~+~
 \left[ 4\hat t (\hat t - m_\mu ^2)\right] 
\left[{ \left\vert g_{2L} \, g_{2R}\right\vert^2 \over (\hat u - M_{ V_{2\mu} ^a}^2)^2 } \right]
\label{wru}
\eea
\begin{figure}[hb]
\vskip -4 cm
\centerline{ 
\epsfxsize=12cm\epsfysize=24cm\epsfbox{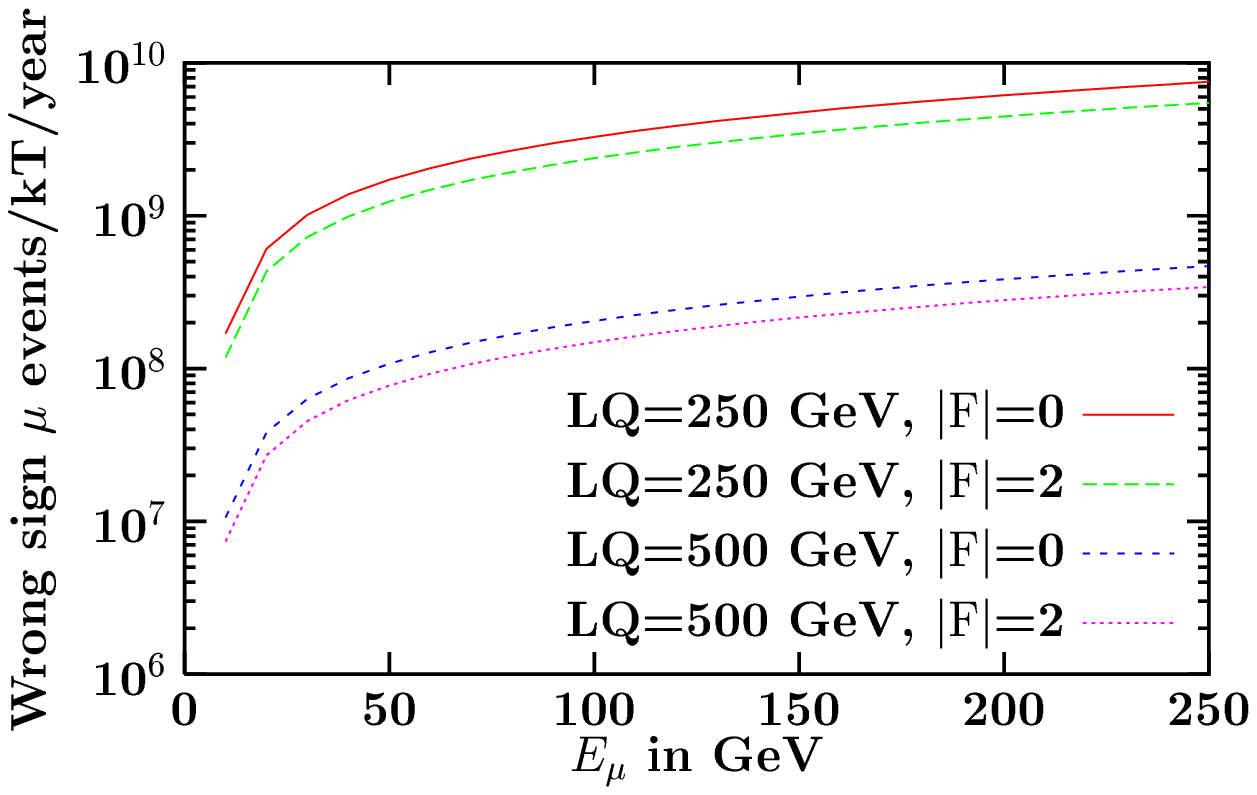}
	\hspace*{-3.5 cm}
\epsfxsize=12cm\epsfysize=24cm
                     \epsfbox{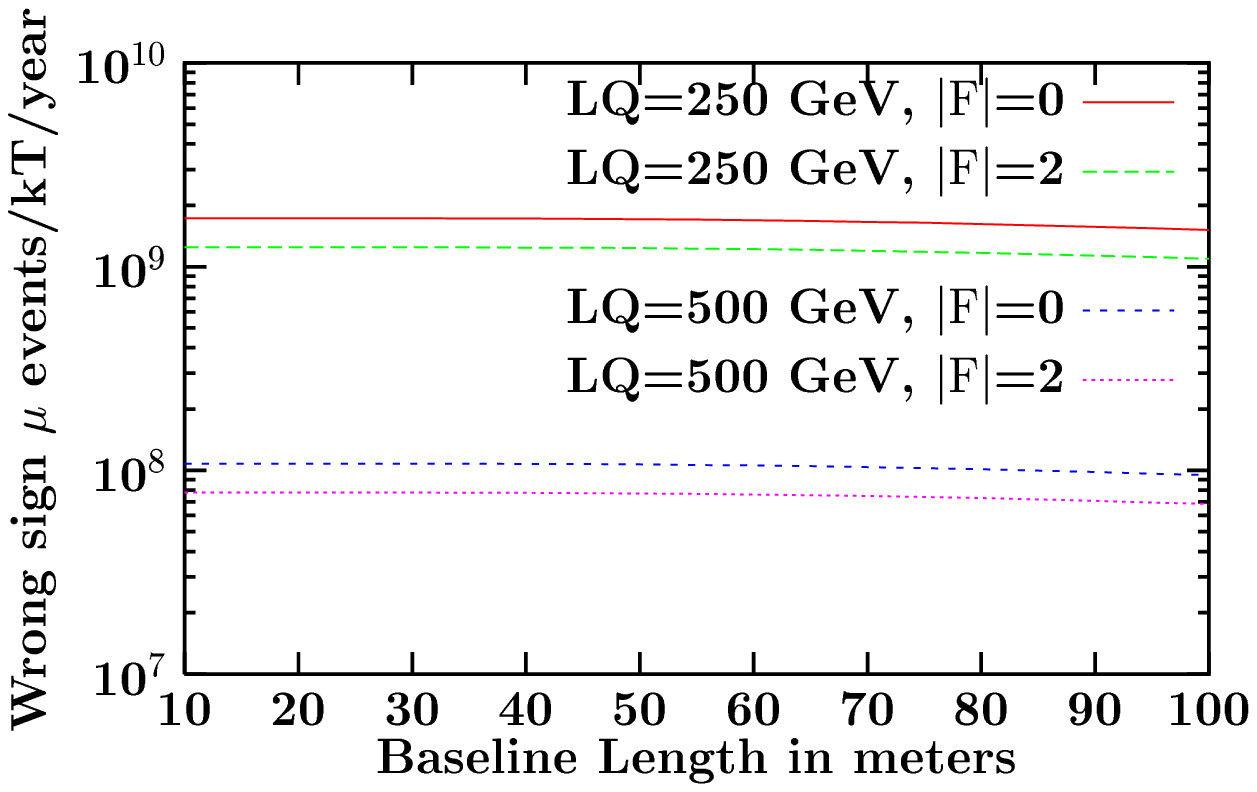}
}
\vskip -13 cm
\hskip 3cm
{\bf (a)}
\hskip 8cm
{\bf (b)}
\vskip -3.5 cm
\centerline{ 
\epsfxsize=12cm\epsfysize=24cm\epsfbox{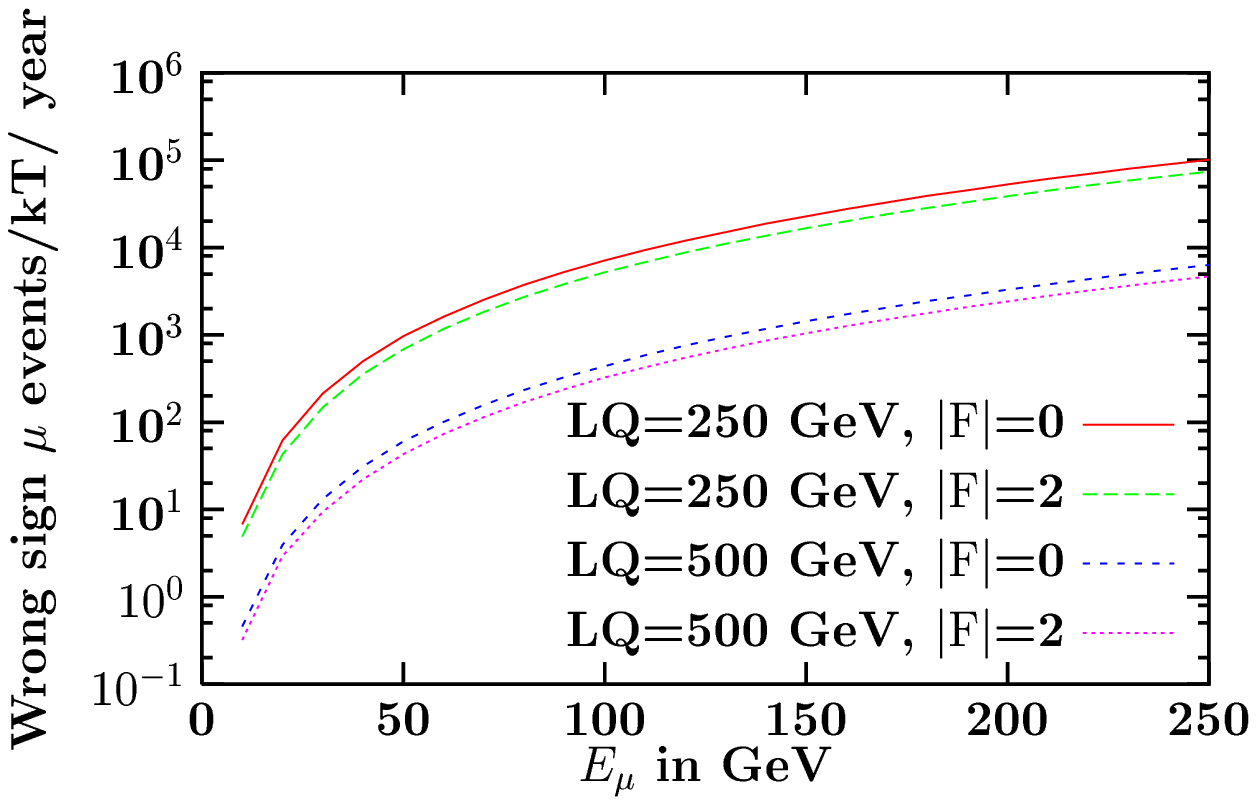}
	\hspace*{-3.5 cm}
\epsfxsize=12cm\epsfysize=24cm
                     \epsfbox{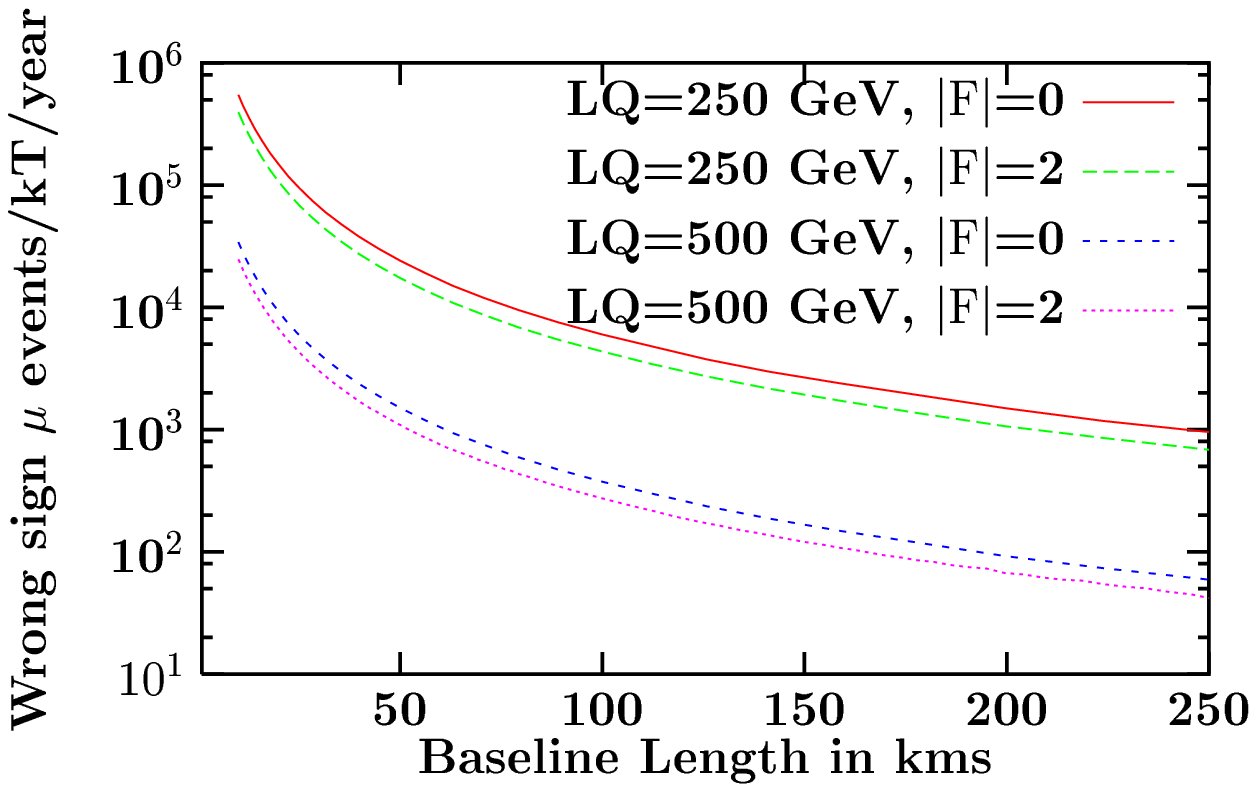}
}
\vskip -13 cm
\hskip 4cm
{\bf (c)}
\hskip 8cm
{\bf (d)}
\caption{{\em Variation of wrong sign $\mu $-events ( from osc. and LQ ) 
with : 
(a) muon beam energy for a baseline length 40 meters, 
(b) baseline length for near-site detector configuration,
(c) muon beam energy for a baseline length 250 kms,
(d) baseline length for short baseline situation. 
All the parameters used here are as mentioned in the caption of figure 2.
}}
\label{fig:fig5}
\end{figure}
Similar expressions of matrix element squared for $s$- and $u$- 
channel 
diagrams corresponding to the process $\bar{\nu_e} u \longrightarrow 
\tau^+ d$ can be obtained just by substituting $m_\mu^2$ by $m_\tau^2$ 
and $p_{\mu^+}$ by $p_{\tau^+}$ in equations (\ref{wrs}) 
and (\ref{wru}).
\par In order to study the behaviour of wrong sign muon events 
w.r.t $E_\mu$ and baseline length, we have used the same 
coupling strengths and masses as mentioned in section 2. 
For the indirect production of $\mu^+$ via decay of $\tau^+$ 
we have taken the  efficiency factor for $\tau$ detection 
(in leptonic channel) to be 30\% \cite{rparity,detector}. 
Predictions for wrong sign muon production rate w.r.t $E_\mu$ 
and baseline length are plotted in figure \ref{fig:fig5}. 
The features of the plots for both near-site and short baseline 
experiments are same as that for $\tau$ production case
discussed in the previous section.
\par In our calculation, 
we have not put any specific selection cut for the 
production of wrong sign $\mu$. 
However, the muons from charm decay which forms a significant 
background for the production of wrong sign muons, can be eliminated 
by incorporating stringent cuts on the transverse momentum of muons, 
missing $p_T$ and isolation cut as mentioned in \cite{rparity,detector,cuts}.
\begin{center}
\begin{figure}[hb]
\input{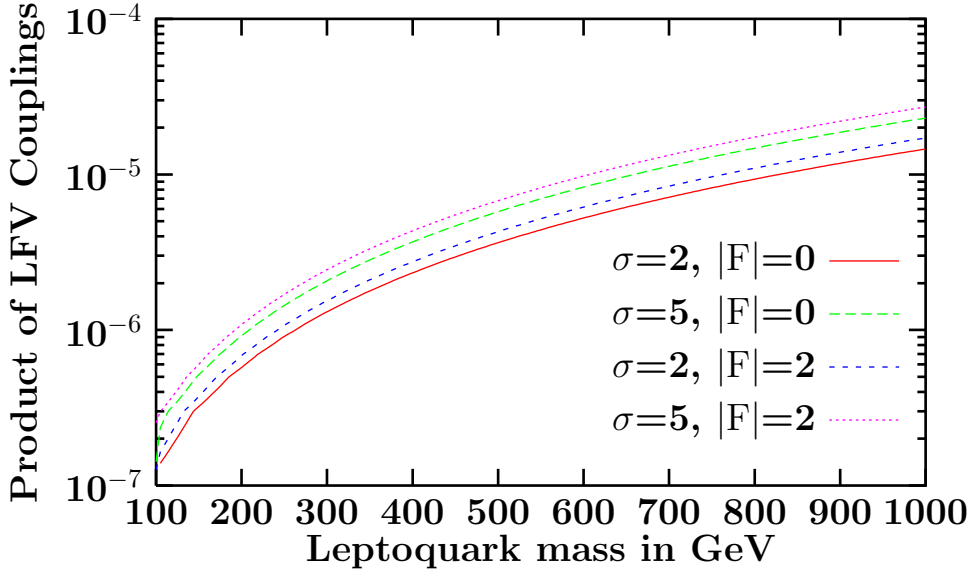}
\vskip .3in
\caption{{\em Contour plot for wrong sign muons at 
2$\sigma$ and 5$\sigma$ effect for $E_\mu$=50 GeV, 
baseline length=40 meters and 
sample detector of area 2500 cm$^2$ and mass 1kT.}}
\label{fig:fig6}
\end{figure}
\end{center} 
\noindent {\bf {Sensitivity Limits \,:}}
Accepting the requirement of $2\sigma$ and $5\sigma$ effect 
as a sensible discovery criterion, we plot the corresponding 
contours for the wrong sign muons at a 
baseline length=40 m in figure \ref{fig:fig6}. 
\end{section}

\begin{section}
{$\tau$ and Wrong Sign $\mu$ Appearance at a NF and Low Energy Bounds}

 In sections 2 and 3 for the purpose of illustration, 
we considered ${\vert {\rm F}\vert}=0$ and 
${\vert {\rm F}\vert}=2$ couplings separately and 
took all couplings to be equal to the 
electromagnetic coupling, $\alpha_{em}$. 
But as also discussed in the introduction, 
strong constraints on the LQ couplings and masses have been 
obtained in the literature from FCNC processes \cite{davidson}. 
In particular, bounds obtained from rare $\tau$ decay 
$\tau \longrightarrow \pi^0 \mu $ and 
from $\mu \longleftrightarrow e$ conversion in nuclei 
would have a direct bearing on the processes considered here. 
This is because low energy limits put stringent bounds 
on effective four-fermion interactions involving two leptons 
and two quarks and since at a NF the centre of 
mass energy in collisions is low enough, we can consider 
the neutrino-quark interactions as four-fermion interactions. 
These bounds on the effective couplings given as LQ couplings 
over mass squared of the LQ are derived on the assumption 
that individual LQ coupling contribution to the branching 
ratio does not exceed the experimental upper limits and 
in the branching ratios only one LQ coupling contribution is 
considered by \lq switching off \rq  all the other couplings. 
The couplings are taken to be real but in these studies 
combinations of left and right chirality couplings are not considered.
\par Based on  these studies, we make some simplified 
assumptions like obtaining the product of couplings of different 
chirality from the square of couplings of individual chirality. 
We extract the coupling products relevant to ($\nu_\mu\, d$) ($\tau\, u$) 
vertex from rare $\tau$ decay bounds as quoted in the 
reference \cite{davidson} and we get the following 
\begin{eqnarray}
\left\vert h_{1L}\right\vert^2=
\left\vert h_{1R}\right\vert^2=
1.9\times 10^{-3}\,\,
\left( {M_{LQ}\over 100\,\, {\rm GeV}}\right)^2,
&& \left\vert h_{2L}\right\vert^2=
3.9\times 10^{-3}\,\,
\left( {M_{LQ}\over 100\,\, {\rm GeV}}\right)^2,
\nonumber\\
\left\vert h_{3L}\right\vert^2=
6.4\times 10^{-4}\,\,
\left( {M_{LQ}\over 100\,\, {\rm GeV}}\right)^2,
&& \left\vert h_{2R}\right\vert^2=
1.9\times 10^{-3}\,\,
\left( {M_{LQ}\over 100\,\, {\rm GeV}}\right)^2,
\nonumber\\
\left\vert g_{1L}\right\vert^2=
\left\vert g_{1R}\right\vert^2=
3.9\times 10^{-3}\,\,
\left( {M_{LQ}\over 100\,\, {\rm GeV}}\right)^2,
&& \left\vert g_{3L}\right\vert^2=
1.3\times 10^{-3}\,\,
\left( {M_{LQ}\over 100\,\, {\rm GeV}}\right)^2,
\nonumber\\
\left\vert g_{2L}\right\vert^2=
1.9\times 10^{-3}\,\,
\left( {M_{LQ}\over 100\,\, {\rm GeV}}\right)^2,
&& \left\vert g_{2R}\right\vert^2=
9.7\times 10^{-4}\,\,
\left( {M_{LQ}\over 100\,\, {\rm GeV}}\right)^2.
\end{eqnarray}

\begin{figure}[h]
\vskip -4cm
\centerline{ 
\epsfxsize=12cm\epsfysize=24cm\epsfbox{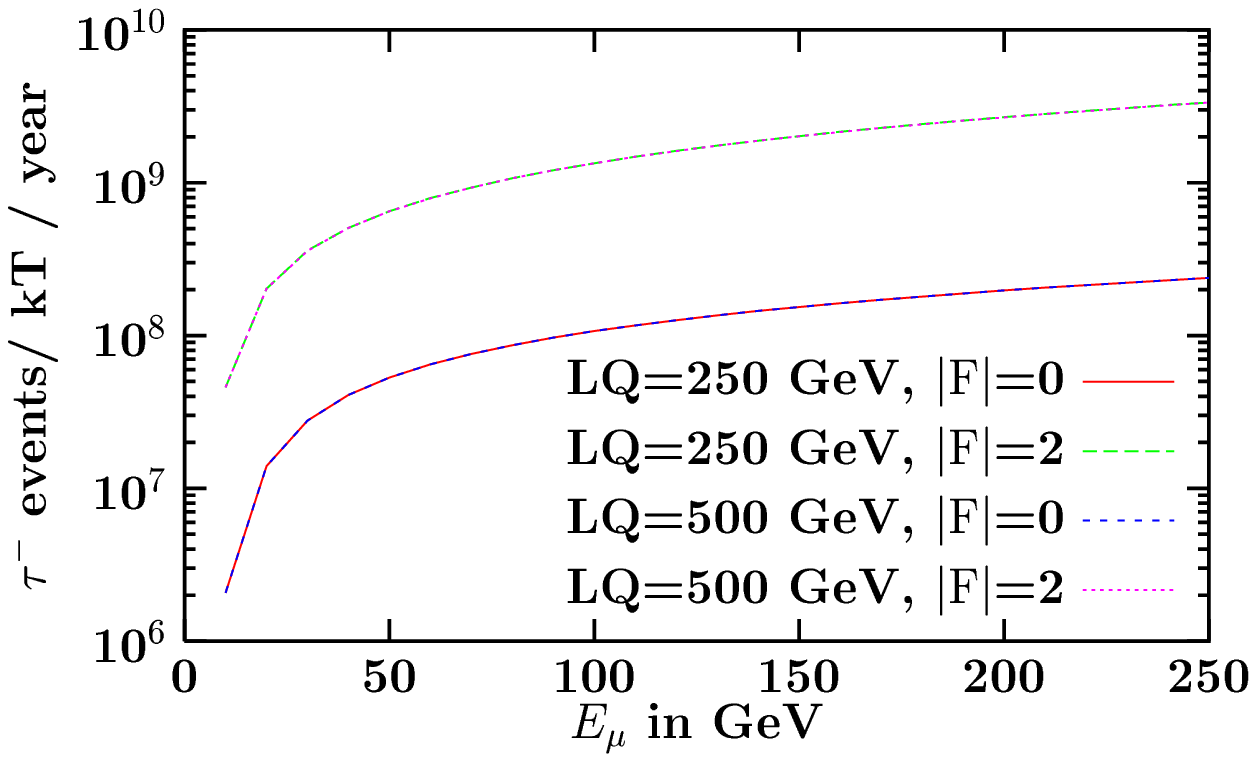}
	\hspace*{-3.5 cm}
\epsfxsize=12cm\epsfysize=24cm
                     \epsfbox{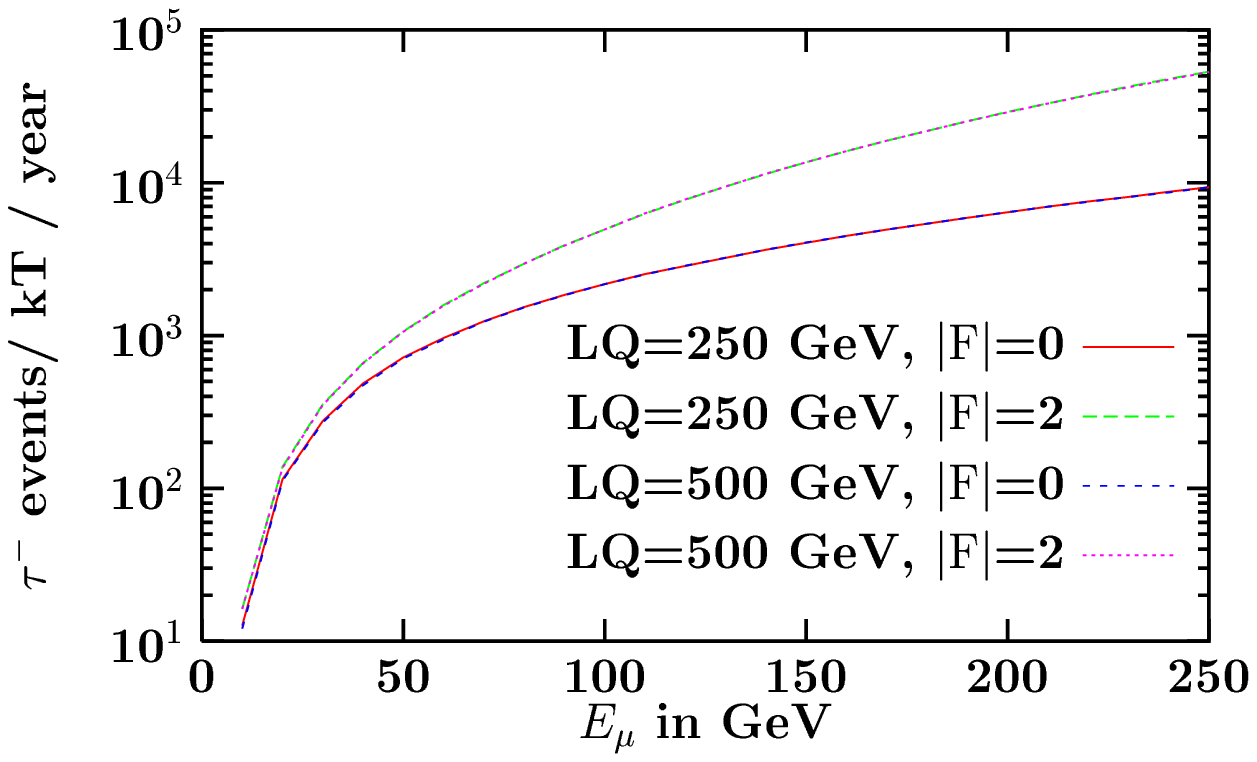}
}
\vskip -13cm
\hskip 4cm
{\bf (a)}
\hskip 8cm
{\bf (b)}
\caption{{\em Variation of  $\tau $-events ( from osc. and LQ ) 
with : (a) muon beam energy for a baseline length 40 meters, 
(b) muon beam energy for a baseline length 250 kms.
All the parameters used here except for the couplings are as 
mentioned in the caption of figure 4.}}
\label{fig:fig7}
\end{figure}
\par In case of wrong sign $\mu$, the bounds on the 
couplings for $(\bar \nu_e u)(\mu^+ d)$ vertex arising from 
$\mu \longleftrightarrow e$ conversion are so stringent, being 
typically 
2-3 orders of magnitude lower compared to bounds on  couplings involving 
third generation of quarks and leptons, that the direct production 
of $\mu^+$ is highly suppressed.
The relevant coupling constants extracted from \cite{davidson}  are
\begin{eqnarray}
\left\vert h_{1L}\right\vert^2=
\left\vert h_{1R}\right\vert^2=
2.6\times 10^{-7}\,\,
\left( {M_{LQ}\over 100\,\, {\rm GeV}}\right)^2,
&& \left\vert h_{2L}\right\vert^2=
5.2\times 10^{-7}\,\,
\left( {M_{LQ}\over 100\,\, {\rm GeV}}\right)^2,
\nonumber\\
\left\vert h_{3L}\right\vert^2=
8.5\times 10^{-8}\,\,
\left( {M_{LQ}\over 100\,\, {\rm GeV}}\right)^2,
&& \left\vert h_{2R}\right\vert^2=
2.6\times 10^{-7}\,\,
\left( {M_{LQ}\over 100\,\, {\rm GeV}}\right)^2,
\nonumber\\
\left\vert g_{1L}\right\vert^2=
\left\vert g_{1R}\right\vert^2=
5.2\times 10^{-7}\,\,
\left( {M_{LQ}\over 100\,\, {\rm GeV}}\right)^2,
&& \left\vert g_{3L}\right\vert^2=
1.7\times 10^{-7}\,\,
\left( {M_{LQ}\over 100\,\, {\rm GeV}}\right)^2,
\nonumber\\
\left\vert g_{2L}\right\vert^2=
2.6\times 10^{-7}\,\,
\left( {M_{LQ}\over 100\,\, {\rm GeV}}\right)^2,
&& \left\vert g_{2R}\right\vert^2=
1.3\times 10^{-7}\,\,
\left( {M_{LQ}\over 100\,\, {\rm GeV}}\right)^2.
\end{eqnarray}
 In this situation, wrong sign muons mainly arise through the 
production of $\tau^+$'s, which subsequently decay via leptonic channel. 
The bounds on coupling constants for the $(\bar \nu_e u) (\tau^+ d)$ 
vertex come from the decay $\tau \longrightarrow \pi^0 e$ and 
are essentially the same as that for the case of $\tau$ 
production \cite{davidson}. 
\begin{figure}[h]
\vskip -4cm
\centerline{ 
\epsfxsize=12cm\epsfysize=24cm\epsfbox{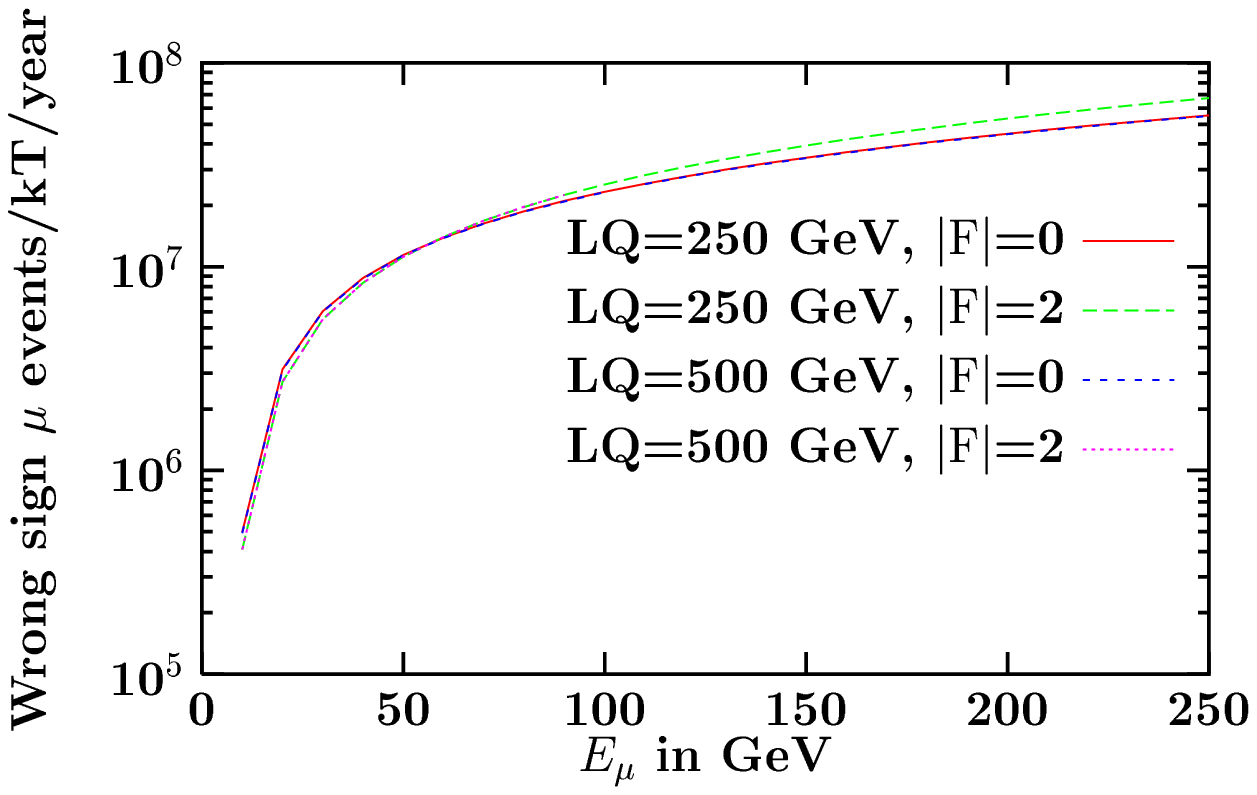}
	\hspace*{-3.5 cm}
\epsfxsize=12cm\epsfysize=24cm
                     \epsfbox{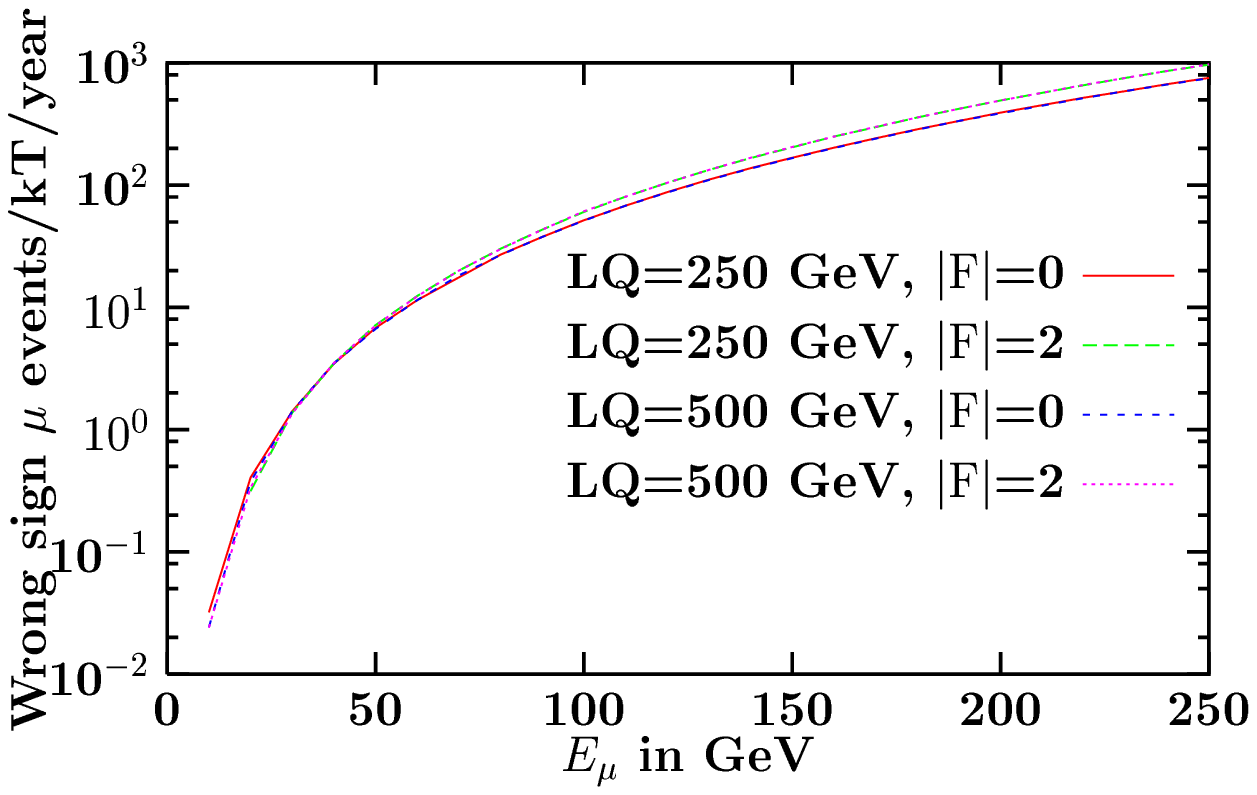}
}
\vskip -13.5cm
\hskip 4cm
{\bf (a)}
\hskip 8cm
{\bf (b)}
\caption{{\em Variation of wrong sign $\mu $-events ( from osc. and LQ ) 
with : 
(a) muon beam energy for a baseline length 40 meters, 
(b) muon beam energy for a baseline length 250 kms.
All parameters used here for plotting except for couplings 
are as quoted in the caption of figure 5.}}
\label{fig:fig8}
\end{figure}
\par In figure \ref{fig:fig7} we show the variation of $\tau $ 
events with muon beam energy and in figure \ref{fig:fig8}, 
the variation of wrong sign muons with muon energy for the baseline 
lengths of 40 m and 250 km respectively. The graphs clearly show 
that number of $\tau$/ wrong sign muons are independent of 
LQ masses, as expected. On comparing figures 7 \& 8 with 
figures 2 \& 5 respectively, we find considerable suppression 
in event rates. 
\par We should however bear in mind that rare decay 
bounds in LQ interactions with a charm quark are comparatively weak and 
therefore these bounds can be evaded if we can tag the charm production.
\end{section}
\begin{section}{Conclusions}
NF will open up unprecendented opportunities to 
investigate $\nu$ physics, bearing not only on $\nu$ oscillation 
phenomenon but also providing physical laboratory for 
testing physics beyond the SM. 
In this letter, we investigated the LFV effect in theories with LQ on the 
production of $\tau $'s and wrong sign $\mu$'s in the near and 
short baseline experiments. 
It is clear that with the increase in baseline length, 
the LQ event rate falls off and neutrino oscillations are the main 
source events examined here. 
At near-site experiments, on the other hand, the events mainly 
arise from {\it new interactions} and can thus be used to constrain 
the theory (figures 3-8). In particular one can obtain 
constraints on LFV couplings between the first and third 
generation, the bounds on which are generally not available.  
At near-site experiments, the event rate is practically 
independent of baseline length.   
\end{section}

\noindent{\bf Acknowledgment:} We are grateful to Namit Mahajan, 
Debajyoti Choudhury and Anindya Datta for useful discussions.
P.M. acknowledges Council for Scientific and Industrial Research, 
India while A.G. acknowledges the University Grants Commission, 
India for partial financial support. 
\newcommand{\plb}[3]{{Phys. Lett.} {\bf B#1,} #2 (#3)}                  %
\newcommand{\prl}[3]{Phys. Rev. Lett. {\bf #1,} #2 (#3)}        %
\newcommand{\rmp}[3]{Rev. Mod.  Phys. {\bf #1,} #2 (#3)}             %
\newcommand{\prep}[3]{Phys. Rep. {\bf #1,} #2 (#3)}                     %
\newcommand{\rpp}[3]{Rep. Prog. Phys. {\bf #1,} #2 (#3)}             %
\newcommand{\prd}[3]{Phys. Rev. {\bf D#1,} #2 (#3)}                    %
\newcommand{\prc}[3]{{Phys. Rev.}{\bf C#1,} #2 (#3)}  
\newcommand{\np}[3]{Nucl. Phys. {\bf B#1,} #2 (#3)}                    %
\newcommand{\npbps}[3]{Nucl. Phys. B (Proc. Suppl.) 
           {\bf #1,} #2 (#3)}                                           %
\newcommand{\sci}[3]{Science {\bf #1,} (#3) #2}                 %
\newcommand{\zp}[3]{Z.~Phys. C{\bf#1,} #2 (#3)}                              
 \newcommand{\ijmpa}[3]{Int.~J. Mod. Phys.{\bf A#1,} #2 (#3)}      
\newcommand{\mpla}[3]{Mod. Phys. Lett. {\bf A#1,} #2 (#3)}             %
\newcommand{\epjc}[3]{Euro. Phys. J.{\bf C#1,} #2 (#3)}
 \newcommand{\apj}[3]{ Astrophys. J.\/ {\bf #1,} #2 (#3)}       %
\newcommand{\astropp}[3]{Astropart. Phys. {\bf #1,} #2 (#3)}            %
\newcommand{\ib}[3]{{ ibid.\/} {\bf #1,} #2 (#3)}                    %
\newcommand{\nat}[3]{Nature (London) {\bf #1,} (#3) #2}         %
 \newcommand{\app}[3]{{ Acta Phys. Polon.   B\/}{\bf #1,} (#3) #2}%
\newcommand{\nuovocim}[3]{Nuovo Cim. {\bf C#1,} (#3) #2}         %
\newcommand{\yadfiz}[4]{Yad. Fiz. {\bf #1,} (#3) #2;             %
Sov. J. Nucl.  Phys. {\bf #1,} #3 (#4)]}               %
\newcommand{\jetp}[6]{{Zh. Eksp. Teor. Fiz.\/} {\bf #1,} (#3) #2;
           {JETP } {\bf #4} (#6) #5}%

\newcommand{\philt}[3]{Phil. Trans. Roy. Soc. London A {\bf #1} #2  
  (#3)}                                                          %
\newcommand{\hepph}[1]{(hep--ph/#1)}           %
\newcommand{\hepex}[1]{ (hep--ex/#1)}           %
\newcommand{\astro}[1]{(astro--ph/#1)}         %

\end{document}